\documentclass{egpubl}
 
\ShortPresentation      
\usepackage[T1]{fontenc}
\usepackage{dfadobe}  

\usepackage{cite}  
\BibtexOrBiblatex
\electronicVersion
\PrintedOrElectronic
\ifpdf \usepackage[pdftex]{graphicx} \pdfcompresslevel=9
\else \usepackage[dvips]{graphicx} \fi

\usepackage{egweblnk} 

\usepackage{pgfplots, pgfplotstable}
\usepackage{subfig}

\usepackage{pbox}

\graphicspath{ {figures/} }

\title[Hash-Based Ray Path Prediction: Skipping BVH Traversal Computation by Exploiting Ray Locality]%
      {Hash-Based Ray Path Prediction: Skipping BVH Traversal Computation by Exploiting Ray Locality}

\author[Francois Demoullin \& Ayub Gubran \& Tor Aamodt]
{\parbox{\textwidth}{\centering Francois Demoullin
        , Ayub Gubran
        and Tor Aamodt
        }
        \\
{
\parbox{\textwidth}{\centering Department of Electrical and Computer Engineering, University of British Columbia} 
}
}

\begin{document}


\maketitle

%
%
%

%
%

\begin{abstract}
State-of-the-art ray tracing techniques operate on hierarchical acceleration structures such as BVH trees which wrap objects in a scene into bounding volumes of decreasing sizes. Acceleration structures reduce the amount of ray-scene intersections that a ray has to perform to find the intersecting object. However, we observe a large amount of redundancy when rays are traversing these acceleration structures. While modern acceleration structures explore the spatial organization of the scene, they neglect similarities between rays that traverse the structures and thereby cause redundant traversals.

This paper provides a limit study of a new promising technique, Hash-Based Ray Path Prediction (HRPP), which exploits the similarity between rays to predict leaf nodes to avoid redundant acceleration structure traversals. Our data shows that acceleration structure traversal consumes a significant proportion of the ray tracing rendering time regardless of the platform or the target image quality. Our study quantifies unused ray locality and evaluates the theoretical potential for improved ray traversal performance for both coherent and seemingly incoherent rays. We show that HRPP is able to skip, on average, 40\% of all hit-all traversal computations.


\end{abstract}

%
%
\section{Introduction}

Ray tracing techniques~\cite{Whitted:1980:IIM:358876.358882} employ hierarchical acceleration structures, such as Bounding Volume Hierarchies (BVH), that capture spatial locality through subdividing scenes into a hierarchy of ever tighter bounding boxes. These acceleration structures, i.e., traversal trees, reduce the subset of a scene that a ray has to intersect. However, reducing ray-scene calculations comes at the cost of additional ray-box intersections that have to precede ray-scene intersection computations. 

Consequently, a trade-off has to be made between the depth and the width of a traversal tree, where the branching factor determines the depth and width of a tree. Wide trees are shallow and able to quickly traverse a ray to a leaf node for ray-scene intersection computations. On the other hand, deep trees need to traverse many interior levels to reach a leaf node but it entails less ray-scene calculations.

This paper proposes and studies the potential of a new technique, hash-based ray path prediction (HRPP), which reduces the cost of traversing deep trees by exploiting ray locality, where rays from close-by origins and similar directions follow a similar path through the tree. HRPP exploits ray locality present throughout a scene traversal to skip redundant ray-box intersections. 

Extensive work has been done in recent years exploiting ray coherence for efficient traversal by mapping coherent rays to parallel hardware such as SIMD units and GPU Warps \cite{doi:10.1111/1467-8659.00508, Wald07simdray, Pharr:1997:RCS:258734.258791}. HRPP is different in that it proposes to skip redundant ray traversal computations and prevent rays with high locality to previous rays from even entering the acceleration structure. Skipping interior node traversal avoids all DRAM traffic associated with interior nodes and reduces to overall computations needed to ray trace a frame.

Similar to the limit study by Lam and Wilson \cite{Lam:1992:LCF:139669.139702} which presents an upper bound on control flow parallelism, we present a limit study on how many interior nodes can be skipped by exploiting ray locality using HRPP.

This paper makes the following contributions:
\begin{itemize}
\item It highlights the potential of ray path prediction for accelerating hierarchical tree traversal;
\item It proposes a hash function that takes into account ray properties so it can be used for predicting the path of similar rays;
\item It shows a theoretical limit study that quantifies ray locality and evaluates its potential for performance improvements using HRPP, which was able to avoid 30\% of ray-box intersections. This limit study serves as a basis of a future hardware-based implementation that can harness HRPP to improve ray tracing performance.
\item It presents an implementation with open-source code at \href{https://bitbucket.org/FrancoisDemoullin/pbrt_francois_copy}{https://bitbucket.org/FrancoisDemoullin/pbrt\_francois\_copy}.
\end{itemize}

%
%
\section{Related Work}

\textit{Ray Locality.} Pharr et al. ~\cite{Pharr:1997:RCS:258734.258791} observed that primary rays and simple light rays exhibit ray locality but chose not to further explore this type of locality. Boulos et al.~\cite{Boulos2008AdaptiveRP} worked on pre-processing rays in the ray pool to accelerate ray traversal, however, their technique is limited to offline rendering and infeasible for online rendering. Wald et al \cite{doi:10.1111/1467-8659.00508, Wald07simdray} use ray locality for packet tracing but do not skip redundant computations.

\textit{GPU BVH Traversal.} Aila and Laine ~\cite{Aila:2009:UER:1572769.1572792} provided the most widely used GPU benchmark for ray traversal. Aila and Kerras ~\cite{Aila:2010:ACT:1921479.1921497} propose to reduce the memory traffic during tree traversal via changes to the GPU architecture to facilitate treelet compaction. Aila and Laine's GPU ray tracing benchmark has been further extended recently by Ylitie et al. ~\cite{Ylitie:2017:EIR:3105762.3105773} where the compression of the memory traffic from an 8-wide BVH improved performance by a factor of up to 3.3 due to memory reduction. Lier et al. ~\cite{Lier:2018:CSR:3231578.3231583} applied SIMD style ray tracing to wide BVH trees to improve incoherent ray traversal performance by 35\%-45\% on average. 

\textit{Harware Accelerated Ray Tracing.} Shkurko et al. ~\cite{Shkurko:2017:DSH:3105762.3105771} achieved performance gains by buffering rays inside the acceleration structure, thereby achieving perfect prefetching and avoiding duplicate memory fetches. Their technique assumes custom MIMD hardware and fixed function logic. More recently, Lu et al. ~\cite{Lu:2017:UPG:3123939.3124532} made changes to the GPU SIMD architecture to minimize thread divergence to improve traversal speed. Thread divergence was avoided by preventing certain threads in warps from intersecting scene data, while others are traversing or fetching new ray data.

\textit{Ray tracing APIs.} Recently ray tracing graphics APIs have enjoyed much progress. Embree ~\cite{Wald:2014:EKF:2601097.2601199} is a highly optimized CPU path tracer, OptiX ~\cite{Parker:2010:OGP:1833349.1778803} is it's GPU counterpart. Microsoft announced the DXR API ~\cite{MicrosoftDXR} which was used recently by EA ~\cite{Haines2019} in a hybrid real-time rendering demo producing great visual accuracy at interactive frame rates.

%
%
\section{Ray Tracing Performance Evaluation}

\begin{figure}[h]
  \centering
  \includegraphics[width=.8\linewidth]{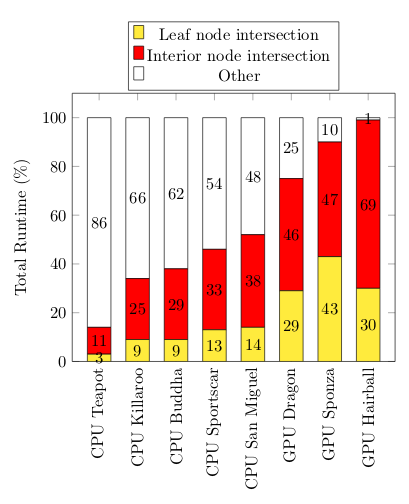}
 
  \caption{\label{fig:figure1}
           Runtime proportions of increasingly complex scenes on CPU and GPU - 1024$\times$1024 - 8 spp}
\end{figure}

We studied the performance characteristics of hierarchical acceleration structures using two BVH-based ray tracing implementations: the CPU-based PBRT~\cite{Pharr:2010:PBR:1854996} renderer and Aila's~\cite{Aila:2009:UER:1572769.1572792} GPU-based renderer. 

PBRT differs from Aila's in that it renders scenes at a higher visual quality. Aila's GPU implementation is optimized for speed and real-time image generation; hence, less effort is put into sophisticated sampling and lighting computations while texture lookups are omitted.

Figure~\ref{fig:figure1} shows the proportion of the total run-time consumed on rendering ray-box intersections (interior nodes) and ray-scene intersections (leaf nodes) in both implementations on increasingly complex scenes. The proportion labeled `Other' in Figure~\ref{fig:figure1} is spent on path tracing tasks such as sampling, texture lookups or lighting computations. The proportion of time spent computing ray-box and ray-triangle intersections is correlated with the geometrical complexity of the scene.

For the CPU-based PBRT, the time spent in traversing the BVH tree is, on average, 40\% of the total execution time, where around 80\% of the traversal time is spent on ray-box intersections.

For Aila's GPU-based ray tracer, the proportion of time spent on BVH traversal is, as expected, higher compared to PBRT. Figure~\ref{fig:figure1} shows that the GPU path tracer is spending upwards of 95\% of the execution time performing tree traversal. Similar to PBRT, the most time-consuming step is ray-box intersections.

The results above show that tree traversal remains the most time-consuming step in ray tracing, even when using benchmarks that use very different rendering implementations and target different levels of image quality. Thus, accelerating the traversal process would benefit both real-time ray tracing (e.g., Aila's GPU implementation) and offline ray tracing (e.g., PBRT), as in both cases calculating ray-box intersections remains the most costly part of the process.

In this paper, we provide a study showing the potential of using Hash-Based Ray Path Prediction to reduce the amount of ray-box intersections calculations (i.e., interior node traversal). Whereby, taking into account the properties of rays and their path through a scene, we can predict which part of a scene a ray intersects; thus, avoiding traversing internal tree levels to reach target leaf nodes.

%
%
\section{Hash-Based Ray Path Prediction}

\begin{figure}[htb]
  \centering
  \includegraphics[width=0.7\linewidth]{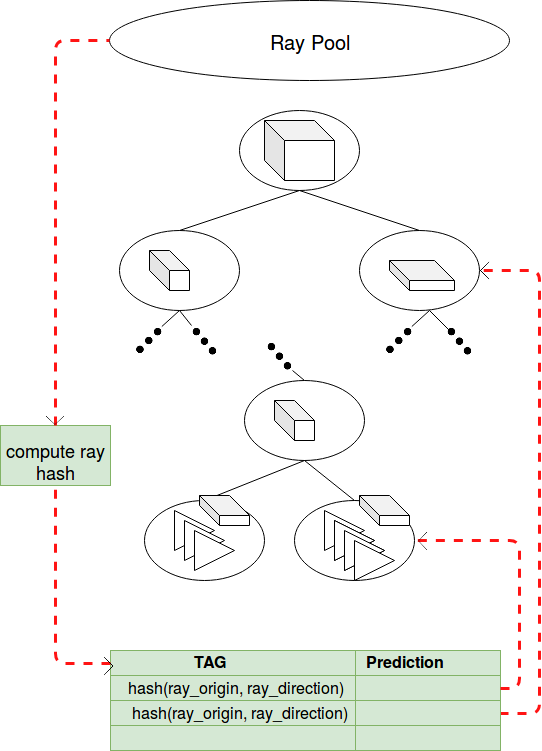}
  %

  %
  %
\caption{\label{fig:idea}
           Binary BVH Acceleration Structure augmented by HRPP. The interior nodes are represented by their bounding boxes. Leaf nodes have both bounding boxes and triangular scene geometry. HRPP additions are marked in green. The red arrows represent the control flow from root to leaves through HRPP}
\end{figure}

\subsection{Overview}
Figure~\ref{fig:idea} shows an example of a binary BVH tree used in PBRT and Aila's GPU raytracer. 

The traversal of rays through the acceleration structure starts at the ray pool which dynamically collects all the rays to be traced throughout the rendering process. Once the rays exit the ray pool, they are traced through the BVH tree in a top-down fashion. On the CPU, ray traversal bundles each group of rays in packets for SIMD parallelized traversal~\cite{Wald07simdray}. On the GPU, rays are bundled into much wider packets called warps, which match the underlying architecture of GPUs~\cite{Aila:2010:ACT:1921479.1921497}.

A ray that intersects with the scene geometry has to traverse up to the full depth of the tree before calculating any scene intersections. Traversing interior-nodes, performing ray-box intersections along the way, is a costly operation that does not provide any merit other than guiding rays to leaf nodes where the essential scene intersections take place.

Shadow Rays follow the `hit-any' paradigm, whereby they search for any scene intersection. Most other rays, however, need to determine the closest intersection to the ray's origin and, therefore, follow the `hit-all' paradigm. 

If two rays exhibit locality they are mapped to similar leaf nodes. Trivially, primary rays all start at the camera and their directions remain similar; consequently, primary rays are highly coherent and exhibit locality with each other. 

Subsequent rays, such as reflections rays, shadow rays, and incoherent global illumination rays also exhibit locality. Many secondary rays, which have originated from distinct primary rays, intersect with similar surfaces in the scene or are traced towards similar light sources. Even seemingly incoherent rays can show similarity with other incoherent rays throughout the rendering of a frame. This type of locality, which goes beyond primary rays, is unexploited in today's raytracers.

\subsection{Hash-Based Ray Path Prediction (HRPP)}
HRPP skips traversal operations by predicting ray paths. HRPP predicts which nodes a ray is likely to intersect based on information gathered from previous similar rays.

HRPP's direct flow guides rays directly from the ray pool towards the leaf nodes as shown in Figure~\ref{fig:idea} by dashed red lines. Rather than letting rays enter the acceleration structure via the ray pool, HRPP computes a ray's hash value. The hash function employed by HRPP uses the ray's physical properties of origin and direction to associate a unique number with each ray. This number serves as a lookup index into HRPP's predictor table. Hash functions are discussed in more detail in Section~\ref{hash}.

Once rays are hashed, the predictor table serves as a storage of mappings from unique ray hash values to node indices which can be used as pointers into the tree. 

Before any ray enters the tree, a ray's hash value is computed, followed by a lookup in the predictor table. If the hash value is present in the table, a prediction for the ray is made. The prediction is then evaluated. If the evaluation results in a valid scene intersection, the traversal of all interior nodes is skipped. Otherwise, the ray is added back into the ray pool and traverses the tree as it would without HRPP.

The predictor predicts either a leaf node or an interior node, as is illustrated in Figure~\ref{fig:idea}. When an interior node is predicted, the ray has to traverse the interior layers preceding the target leaf node.

\subsection{Precision and Recall}
As stated in the previous section, HRPP prediction can either hit or miss the target leaf node. We intuitively define HRPP hits as positive predictions, and HRPP misses as negative predictions. We further differentiate between four scenarios, true positives, false positives, true negatives, and false negatives. We will discuss how each of these scenarios can happen and how to deal with them to guarantee correctness.

True positives occur when a ray hits in the predictor and the prediction is deemed correct after evaluation. A prediction is determined to be correct when a ray-scene intersection is found in the predicted leaf node.  This is the best-case outcome, whereby the traversal of the tree is successfully skipped. It is to be noted, however, that there is a slim possibility of a true positive leading to the wrong visual output, as in the case where the ray is required to find the closest intersection, e.g the `hit-all' paradigm. Prediction can cause a ray to not find the closest intersection. This problem is inexistent for `hit-any' rays such as shadow rays. 

On the other hand, false positives occur when a ray hits in the predictor table but the prediction made by HRPP is incorrect. Upon a predictor hit, the ray is tested against the geometry in the predicted node(s). If the ray misses in the node(s), the prediction is incorrect. In this case, the ray is added back into the ray pool and traverses the acceleration structure as it would without HRPP. The performance cost of a false positive is the cost of the lookup in HRPP and the cost of intersecting the triangles in the wrongly predicted leaf nodes. 

True negatives and false negatives are similar to each other in the way that they simply miss in the predictor but they do so for different reasons. The reason for the miss for a true negative is that there was no chance a similar ray has been encountered previously. False misses occur when an entry has been evicted from the predictor. These misses could have been avoided by using an infinitely sized predictor table. Our current implementation lacks a replacement policy as we focus on evaluating the limits of our technique. We will therefore not differentiate between true negatives and false negatives. 

The penalty for negatives is the cost of the predictor table lookup which includes the cost of computing the hash function. Once a ray has been established as being falsely predicted, it is re-added to the ray pool awaiting traversal through the tree.

%
%

\section{HRPP Hash Function} \label{hash}
Hashing is used to map ray properties to a unique value that serves as an index into the predictor table. In our case, a good hash function maps rays with similar properties to the same entry in the predictor table.

\begin{figure}[htb]
  \centering
  \includegraphics[width=1.0\linewidth]{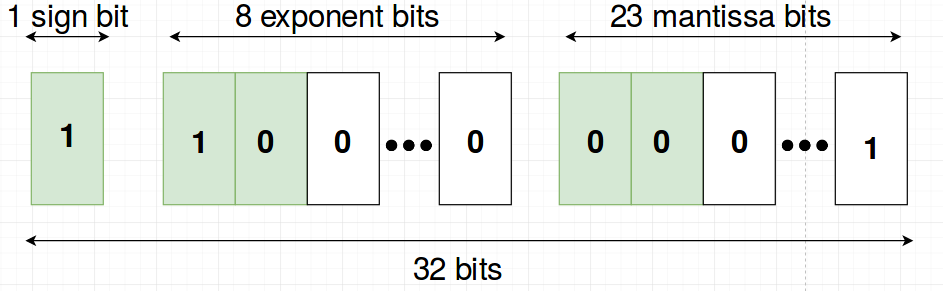}
 
  \caption{\label{fig:IEEEmapping}
           Illustration of our hash function's extraction of bits from IEEE 754 floating point with an example precision of 2 bits. Bits marked in green are included in the hash representation}
\end{figure}

Most path tracers encode ray origin and direction as single precision floating points encoded in the IEEE 754~\cite{1985--ieee754} floating point representation. HRPP's hash function extracts relevant bits from the IEEE754 floating point encoding to create a relevant hash index. 
Furthermore, we propose to implement the hash encoder using a fixed function hardware block for maximal performance. Figure~\ref{fig:IEEEmapping} illustrates the proposed hash encoding using an example with a hash precision of 2 bits. In this example, the predictor index is composed of the sign bit and the 2 most significant bits from the exponent as well as the mantissa. A hash precision of 3 bits would be composed of the sign bit and 3 bits from each exponent and mantissa. 

The same hash encoding is used for all 6 floats of each ray, where 3 floats are used for direction and 3 floats are used for origin. Finally, we swizzle origin hash values with direction hash values to obtain a unique predictor table entry that fits into at most 48 bits. The pseudo-code for the hash function is as follows:

\begin{verbatim}
// extract up to 16 bits from IEEE float representation 
// uint8_t may overflow based on hash precision, use uint16_t for hash \
// precision upwards of 4 bits
uint8_t hash_o_x = map_float_to_hash(ray.origin.x)
/* repeat for all 6 floats */

// xor the hashes to save space
uint8_t hash_0 = hash_o_x xor hash_d_z;
uint8_t hash_1 = hash_o_y xor hash_d_y;
uint8_t hash_2 = hash_o_z xor hash_d_x;

// form a unique index
unsigned long long predictor_table_index = \
    (hash_0 << 0) or \
    (hash_1 << 8) or \ 
    (hash_2 << 16);
\end{verbatim}

While this encoding minimizes hash conflicts, it deliberately does not avoid them; we want similar rays to result in hash conflicts by design.
When encountering a hash conflict, the leaf node of the conflicting ray is inserted into the existing predictor entry. No predictor entry can hold duplicate nodes; therefore, the leaf is only inserted if it is not already present in the predictor entry. 

\subsection{Hash function tradeoff}
In this section, we discuss the tradeoff between using loose vs. tight hash functions.

We define a hash function as loose if it maps too many rays to the same hash index. As a result, the number of false positives increases and the overall number of skipped computations decreases. A symptom of this is that the entries in the predictor table become long because the predictor table associates many distinct leaves with each hash index. On a predictor hit, a ray has to check against all nodes in the table entry. Long table entries result in a large amount of predictor induced overhead computation and few skipped nodes. 

On the other hand, a tight hash function will only map rays to the same index if they are very similar. When using a tight hash function, HRPP uses significantly more memory due to a larger number of distinct hash values being stored in the predictor table. Tight hashing produces few positives and, as a result, few computations are skipped. The upside to a tight hash function, however, is that almost all positives are true positives.  

%
%

\begin{table*}[t]
\centering
\begin{tabular}{ |c||c|c|c|c|c|  }
 \hline
 Scene & Teapot & Killeroo Simple & Killeroo Metal & Buddha & Sportscar\\
 \hline
 number of triangles              & 2k & 60k & 500k & 1mio & 53mio\\
 lighting complexity              & simple & simple & complex & medium & complex\\
 max BVH depth                    & 16 & 24 & 27 & 30 & 32\\
 hit-any savings(\%)              & 4 & 48 & \textless1 & 23 & \textless1\\
 hit-any predictor table size (MB)& 16 & 10.8 & 31 & 18 & 21\\
 hit-all savings(\%)              & 69 & 52 & 30 & 41 & 32\\
 hit-all predictor table size (MB)& 2 & 84 & 47 & 18 & 18\\
 \hline
\end{tabular}
\caption{Evaluation of test scenes - resolution 1024$\times$1024 - 8 spp - Go Up Level 0 - hash precision 6}
\label{tab:table_label}
\end{table*}

\begin{figure}[htb]
  \centering
  \subfloat[]{\includegraphics[width=.8\linewidth]{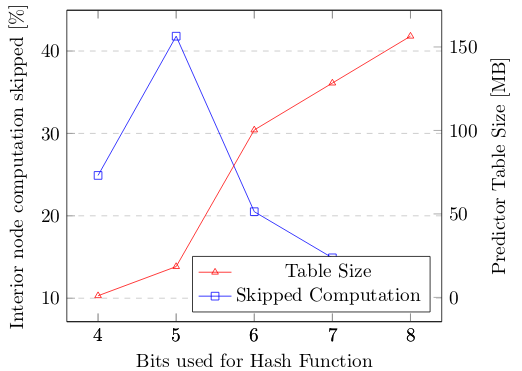}} 
  \newline
  \subfloat[]{\includegraphics[width=.8\linewidth]{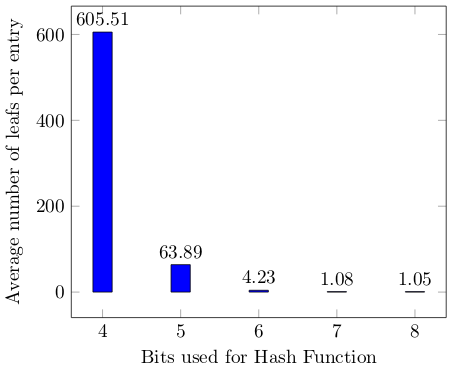}}
 
  \caption{\label{fig:buddha}
           Illustration of the tradeoff between a tight and a loose hash function and its impact on the predictor table properties. Buddha - 1024$\times$1024 - spp 8}
\end{figure}

Figure~\ref{fig:buddha} illustrates this tradeoff using the Buddha scene using PBRT with 8 samples per pixel at a resolution of 1024$\times$1024. The number of precision bits used in the hash function and the size of the predictor table is directly correlated as shown in Figure~\ref{fig:buddha} (a). In this particular example, the number of intersections skipped peaks at 5 bits.

A hash function that is too tight (e.g. 6 bits and up) increases the size of the predictor table without additional computations skipped. 

A hash function that is too loose (e.g 4 bits and below) produces a large number of hash conflicts. This is illustrated in Figure~\ref{fig:buddha} (b), where the average number of leaves per entry significantly increases to over 600 with a precision of 4 bits.

%
%
\section{Limit Study} \label{res}

We implement a software version of HRPP as an extension of the PBRT renderer to quantify the ray locality present in various example scenes. We present an extensive discussion of which parameters are relevant to HRPP and quantify their effect on prediction accuracy. 
 
While previous limit studies, such as the one by Lam and Wilson \cite{Lam:1992:LCF:139669.139702}, show the potential for improvement over the state of the art by quantifying the presence of a new type of locality, they do not provide a practical implementation. This paper, however, presents both a limit study similar to \cite{Lam:1992:LCF:139669.139702} and HRPP as a technique that can be adopted in hardware to produce speedup over the current state of the art raytracers.

For this limit study, our software implementation has the drawback of using a large amount of additional memory, often exceeding over 100MB. A future hardware implementation of HRPP can address the memory problem by implementing a replacement policy for the HRPP predictor table. Future work will study the impact of a replacement policy on the accuracy and the memory consumption of HRPP.

\subsection{Overview Table}
Results are summarized in Table~\ref{tab:table_label}. We evaluate HRPP on five scenes with unified binary BVH trees of increasing geometrical and illumination complexities. We evaluate two distinct HRPP predictors for hit-any and the hit-all rays to avoid hash conflicts between hit-all and hit-any rays. 

We find that the average hit-all prediction ratio is above 30\% across all test scenes. We conclude that hit-all rays are highly suitable for HRPP prediction. 

We find that for scenes with complex lighting models such as the metal killaroo scene and the sports car scene, hit-any HRPP achieves a less than 1\% prediction ratio due to the low locality of global illumination rays. However, on the geometrically complex buddha scene, we skip 30\% of all hit-any rays. We conclude that hit-any rays exhibit little ray locality in scenes with complex lighting models while exhibiting significant ray locality if the lighting model is simple.

\subsection{Go Up Level}

\begin{figure}[h]
  \centering
  \includegraphics[width=.8\linewidth]{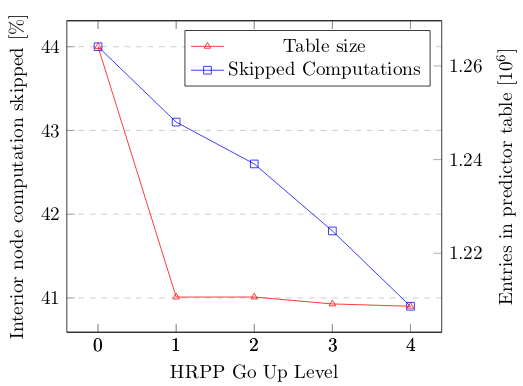}
 
  \caption{\label{fig:buddha_up} Impact of Go Up Level on number of predictor entries. Buddha - 1024$\times$1024 - hash precision 6 }
\end{figure}

We define HRPP's \textit{go up level} as the level in the acceleration structure tree the predictor table predicts. 
A \textit{Go Up Level} of 0 predicts the acceleration structure's leaf nodes. A \textit{Go Up Level} of 1 predicts the parent node of the leaf nodes. A \textit{Go Up Level} of 2 predicts the grand-parent node of the leaf nodes, etc.

As shown in Figure~\ref{fig:buddha_up}, the most significant impact of the predictor's \textit{Go Up Level} is the amount of memory used for the predictor table. A \textit{Go Up Level} of 1 results in fewer entries and therefore less memory used at the cost of a small reduction in skipped computations as compared to a \textit{Go Up Level} of 0.

\subsection{Samples Per Pixel}

\begin{figure}[h]
  \centering
  \includegraphics[width=.8\linewidth]{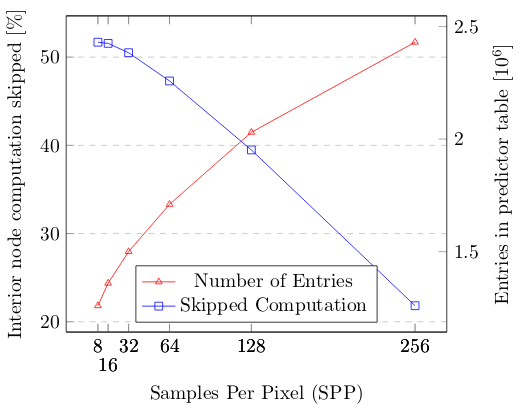}
 
  \caption{\label{fig:spp_result} Impact of SPP on interor node computation skipped and the number of predictor entries. Buddha - 1024$\times$1024 - hash precision 6}
\end{figure}

We study the impact of the number of Samples Per Pixel (SPP) on the efficiency of HRPP in Figure~\ref{fig:spp_result}. 

Counterintuitively, the number of SPPs is inversely correlated with the efficiency of HRPP because of hash conflicts. As the number of hash conflicts increases, the number of memory used for HRPP increases as shown in Figure~\ref{fig:spp_result}. We expect a replacement policy for HRPP to counter this trend but leave the evaluation of replacement policies as future work.

%
%
\section{Conclusion}
This paper explores the benefits of exploiting ray locality to accelerate BVH traversal by skipping interior nodes. In current acceleration structures ray locality goes unutilized, we show that there is speed up to be gained from exploiting this locality. 

Our limit study uncovers significant potential for speed up by skipping on average 30\% of all hit-all rays during the rendering of one frame. 
We explore the design space of ray prediction by quantifying the impact of hash precision, scene complexity, illumination complexity, samples per pixels as well as HRPP's \textit{Go Up Level}. We propose a hash mapping from IEEE 754 floating point and explore the tradeoffs for efficient hashing. 

Finally, this paper provides directions for future work that addresses the current shortcomings of HRPP such as memory consumption and hash conflicts.

%
%



\bibliographystyle{eg-alpha-doi} 
\bibliography{egbibsample}       



\end{document}